\documentclass[prl,twocolumn]{revtex4-1}
\usepackage{bm}
\usepackage{epsfig}
\usepackage{amsmath}
\newcommand{\be}{\begin{equation}}
\newcommand{\ee}{\end{equation}}
\newcommand{\ba}{\begin{eqnarray}}
\newcommand{\ea}{\end{eqnarray}}
\newcommand{\bml}{\begin{mathletters}}
\newcommand{\eml}{\end{mathletters}}
\newcommand{\bes}{\begin{subequations}}
\newcommand{\ees}{\end{subequations}}

\newcommand{\bi}{\begin{itemize}}
\newcommand{\ei}{\end{itemize}}

\begin{document}
\title{Implication of a Quasi Fixed Point with a Heavy Fourth Generation: The emergence of a TeV-scale physical cutoff}
\author{P.Q. Hung}
\email[]{pqh@virginia.edu} 
\author{Chi Xiong}
\email[]{xiong@virginia.edu}
\affiliation{Dept. of Physics, University of Virginia, \\
382 McCormick Road, P. O. Box 400714, Charlottesville, Virginia 22904-4714,
USA}

\date{\today}
\begin{abstract}
It has been shown in a recent paper that the Higgs quartic and Yukawa sectors of the Standard Model (SM) with a heavy
fourth generation exhibit at a two-loop level a quasi fixed point structure instead of the
one-loop Landau singularity  and which
could be located in the TeV region, a scale which is denoted by $\Lambda_{FP} $ in this paper.  This provides the possibility of the existence of a TeV-scale physical cutoff endowed with several implications. In the vicinity of this quasi fixed point
 bound states and Higgs-like condensates made up
of the 4th generation quarks and leptons get formed. 
It implies the possibility of a dynamical electroweak
symmetry breaking generated by 4th generation condensates.  The quasi fixed points also hint at
at a possible restoration of scale symmetry at  $\Lambda_{FP} $  and above and the emergence of 
a theory which could be deeper than the SM.
\end{abstract}

\maketitle

The Standard Model, with all of its successes, has some theoretical shortcomings which hamper its status as a fundamental
theory. Perhaps the most serious issue
with the SM is the existence of a fundamental scalar with its associated quadratic mass divergence problem, especially if the physical cutoff scale is the Planck mass. It is therefore a natural question to ask whether or not one can find a way to lower the physical cutoff to around the electroweak scale.

It has been argued
that a theory is natural if it is stable under tiny variations of fundamental parameters . 
Mass corrections to the fundamental scalar such as the SM Higgs field
are proportional to the physical cutoff if it exists. If the Planck mass is the physical cutoff,
a fine tuning to one part in $10^{38}$ in the coupling is needed if one were to keep the Higgs mass at the electroweak scale. 
It is fair to say that
this line of reasoning has led to important developments in supersymmetric, technicolor, extra dimensional and little Higgs models
\cite{models}. 
From hereon, the term ``hierarchy problem'' will simply refer to the existence
of two scales: the electroweak and Planck scales, and the aforementioned issue. This problem exists {\em regardless}
of whether or not the SM is embedded into some grand unified theory. 

Another line of thought is related to the possibility of a restoration of scale symmetry above a certain energy scale which could be taken as a physical cutoff scale \cite{shapos}.  Notice that, at the quantum level, the trace of the energy momentum tensor
is proportional to the $\beta$ function, i.e. $T^{\mu}_{\mu} \propto \beta (g)$, which indicates the
breaking of scale invariance if $\beta(g) \neq 0$ even if there is scale symmetry at tree level. However, if the theory has a fixed point i.e. $\beta(g) = 0$ at some energy scale $\Lambda_{FP}$, that scale could be taken as a physical cutoff. Scale symmetry is restored at that energy. The physics at or below the physical cutoff scale will be {\em insensitive} to that above $\Lambda_{FP}$. The hierarchy issue might be be ``resolved" if the energy scale where the fixed point is located is in the TeV region. In addition, this possibility could be further strengthened if the electroweak symmetry itself can be dynamically broken close to the fixed point.

\begin{figure}
\includegraphics[angle=0,width=8.5cm]{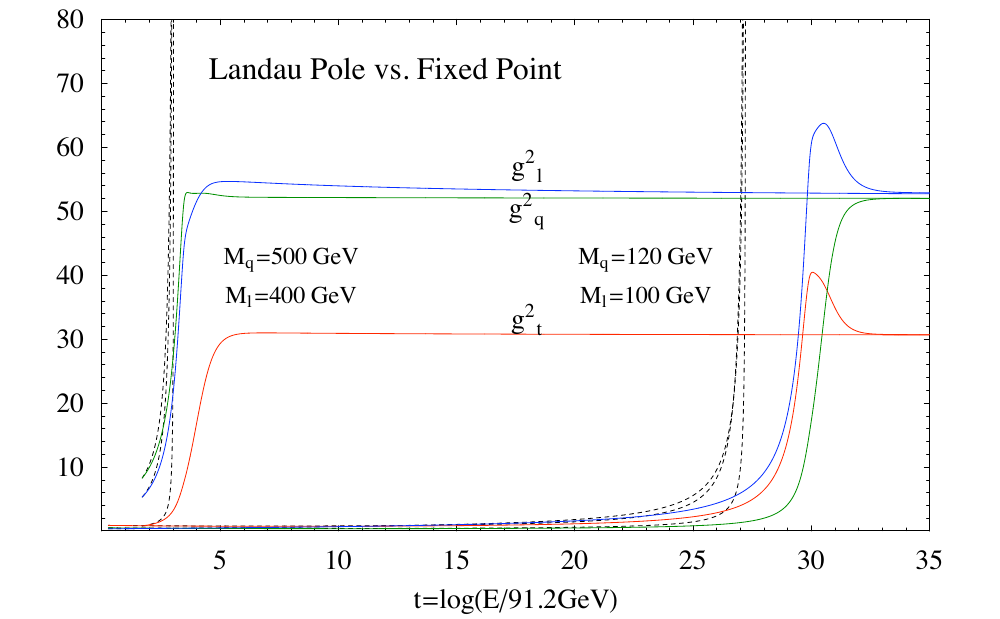}
\caption{{\small The Landau pole(dotted lines) and the quasi fixed point(solid lines) of the Yukawa couplings of the
fourth generation fermions and the top quark. For a heavy fourth generation (left side), both the Landau singularity from
one-loop RGEs  and the quasi fixed point from two-loop RGEs appear at about 2 $\sim$ 3 TeV, while for a light fourth 
generation (right side), their locations at the energy scale differ by two orders of magnitude.  
}}
\label{fig0}
\end{figure}

In this manuscript, we would like to
suggest that the existence of a quasi fixed point \cite{pqchi1}
in the quartic and Yukawa sectors of the SM with a heavy fourth generation provides a natural candidate for a physical cutoff $\Lambda_{FP}$
which could be located in the TeV region.  In fact, this quasi fixed point which appears at two loops in the $\beta$ functions of the Higgs-Yukawa sector provides a {\em solution} to the Landau pole problem which shows up at one loop.  As shown in \cite{pqchi1}, the one-loop Landau singularity appears at approximately the same energy scale as the quasi fixed point when the fourth generation is sufficiently heavy (see Fig.\ref{fig0}). 
The transformation of the Landau pole into a quasi fixed point at a similar scale has important implications: (1) It provides the appearance of a TeV-scale physical cutoff $\Lambda_{FP}$; (2) It provides the possibility of an interesting theory beyond the SM  which lies above the physical cutoff coming from the possibility of the restoration of scale symmetry. 
Furthermore, near this  ``cutoff", the Yukawa
couplings of the 4th generation are large enough for condensates to form and spontaneously
break the electroweak symmetry. Such ultraviolet fixed points in the Higgs-Yukawa sector were previously
found and studied in the context of SU(5) gauge coupling unification \cite{Hung:1997rw} with values comparable to those found
in \cite{pqchi1}. 


Although our results  are based on the existence of a quasi fixed point at the two-loop level, they provide a strong hint at a possible approach to the hierarchy problem. In particular, it is also possible that the $\beta$ functions in the gauge sector might exhibit a quasi fixed point  at the two-loop level thus providing an additional argument in favor of our suggestion. Such gauge quasi fixed had been contemplated some time ago, along with its implications, in a context  in which the SM merges into a larger scale-invariant gauge group and that scale symmetry is broken spontaneously  \cite{hunggen}.

For heuristic purpose, we first present a simplified discussion of the dependence of the scalar mass on the physical cutoff, the so-called quadratic divergence.
A very nice way to treat this ``quadratic divergence''
is to use the intuitive approach of Wilson \cite{wilson} where one divides the momentum integration into ``slices''.  Let us
suppose that there is a physical cutoff scale which we will denote by $\Lambda_{max}$. 
As mentioned above, in \cite{pqchi1} we showed the evolution of the couplings at one and two loops (\ref{fig0}) and one can see that, for a heavy 4th generation, the one-loop Landau singularity appears at a
scale similar to that of the two-loop quasi fixed point. This is the scale that we referred to as  $\Lambda_{FP}$ which will be identified with $\Lambda_{max}$ in this manuscript.

Let us now divide the
separation between the electroweak scale $\Lambda_{EW}$ and the cutoff scale $\Lambda_{max}$ into $n$ equal
slices, each of size $\delta q$, i.e. $\Lambda_{EW}+ n\,\delta q=\Lambda_{max}$, with $\delta q \ll \Lambda_{EW}$. 
When $n \rightarrow \infty$, one of course recovers $\Lambda_{max} \rightarrow \infty$.
Each momentum slice  will be characterized by a ``constant'' value (within that slice) of the (Yukawa or quartic) coupling.
The Wilsonian way to look at the loop integration is to consider the contribution from an $n$ set of theories (for $n$ slices), 
each characterized by the same Lagrangian but endowed with a different coupling constant. 
(The usual divergence encountered in field theory with an infinite cutoff can be viewed as coming from 
the contribution of an {\em infinite} set of such theories.) Let us first look at the fermion loop correction to the scalar mass. 
Schematically, one writes, with $\delta m_{H}^{2} = m_{H}^{2}(\Lambda_{EW})- m_{H,0}^{2}$,
\ba
\label{fermion}
\delta m_{H}^{2} &\approx&
c\,g_{Y}^{2}(\Lambda_{EW})\int_{\Lambda_{EW}^2}^{(\Lambda_{EW}+\delta q)^2} dk^2+\cdots + \nonumber \\
&& 
c\,g_{Y}^{2}(\Lambda_{EW}+(n-1)\,\delta q)\int_{(\Lambda_{EW}+(n-1)\delta q)^2}^{\Lambda_{max}^2} dk^2\,, \nonumber \\
\ea
where $m_{H,0}$ is the tree-level value of the scalar mass  of $O(\Lambda_{EW})$ and $c \sim O(1/16\pi^2)$. A few words are in order at this point.  Eq. (\ref{fermion}) can be transformed into a Renormalization Group equation relating $\delta m_{H}^{2} $ at one energy scale to another at a different scale. It involves the {\em running} of the dimensionless coupling $g_{Y}^{2}$ as one can explicitly see in (\ref{fermion}).  Notice that the usual discussion of the quadratic divergence assumes a constant coupling e.g. a constant Yukawa coupling. The behavior of $g_{Y}^{2}$ can greatly influence the value of 
$\delta m_{H}^{2} $. Below we will present the importance of the TeV-scale physical cutoff and the existence of a quasi fixed point as opposed to a Landau pole.

\begin{figure}
\includegraphics[angle=0,width=8.5cm]{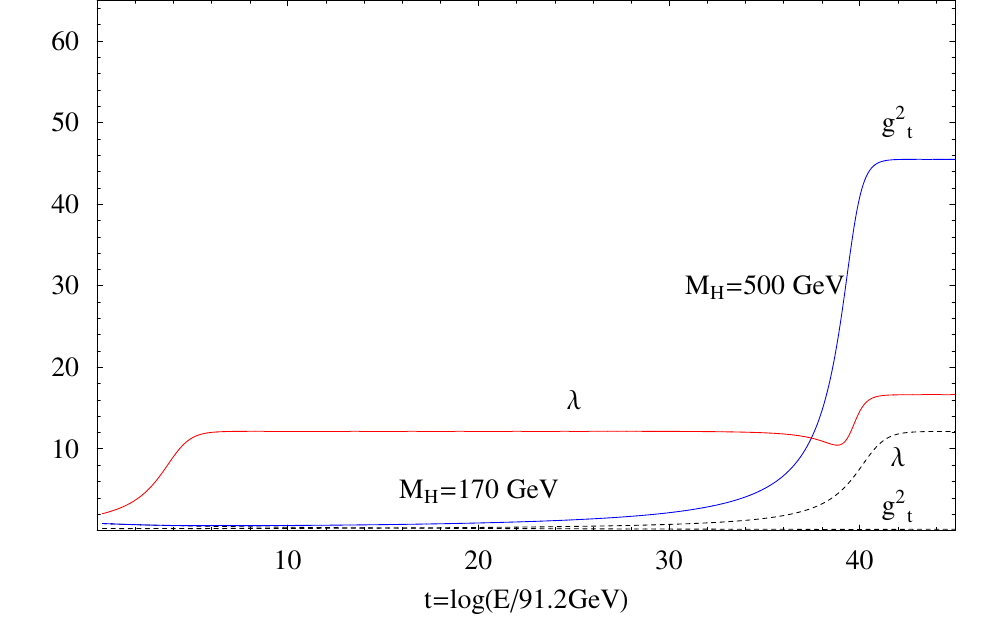}
\caption{The Higgs quartic and top Yukawa couplings at two loops as a function of energy for two
initial Higgs masses. Dashed line: 170 GeV; Solid line: 500 GeV}
\label{fig1}
\end{figure}

\begin{figure}
\includegraphics[angle=0,width=8.5cm]{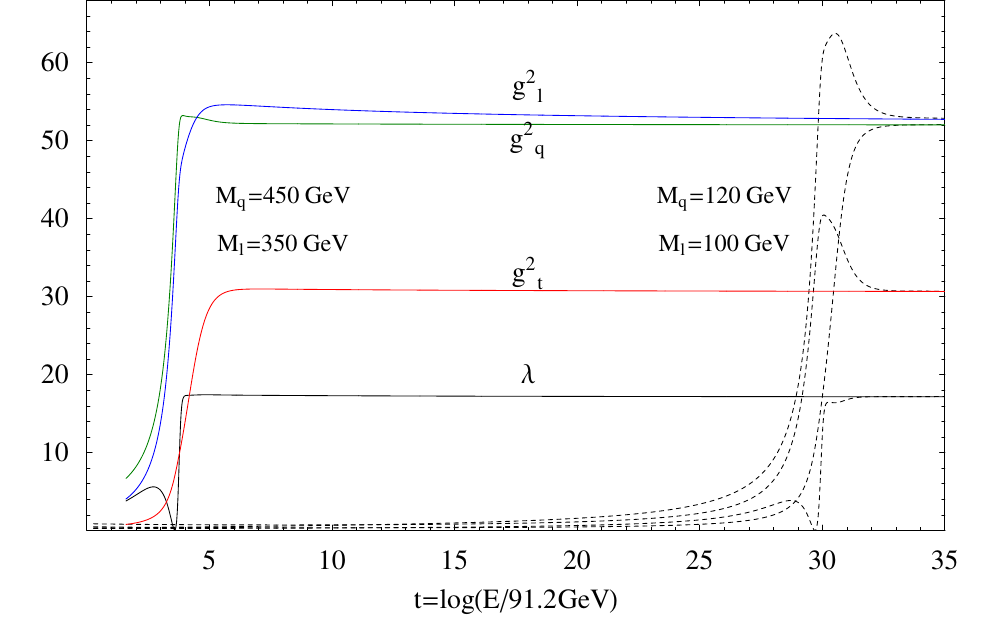}
\caption{The Higgs quartic and 4th generation and top Yukawa couplings at two loops as a function of energy for two
initial Higgs masses}
\label{fig2}
\end{figure}

For the sake of argument, let us first assume that that $g_{Y}^{2}$ is {\em slowly varying}
between $\Lambda_{EW}$ and $\Lambda_{EW}+(n-1)\delta q$. (See Figs. \ref{fig1}, \ref{fig2}.) It
means that $g_{Y}^{2}(\Lambda_{EW})$ is not too different from $g_{Y}^{2}(\Lambda_{EW}+(n-1)\delta q)$. There will not
be a gross error by making the approximation $g_{Y}^{2}(\Lambda_{EW}+(n-1)\delta q)\approx \cdots \approx g_{Y}^{2}(\Lambda_{EW})$.
With that approximation and using $\delta q \ll \Lambda_{EW}$, the mass correction from the fermion loop is approximately
given by
\ba
\label{fermion2}
 \delta m_{H}^{2} &\approx&  c \,g_{Y}^{2}(\Lambda_{EW})\,(2 \Lambda_{EW} \,n \delta q + (n^2-2n)(\delta q)^2) 
\nonumber \\
&\approx& c \,g_{Y}^{2}(\Lambda_{EW})~ \Lambda_{max}^2(1-\frac{\Lambda_{EW}}{\Lambda_{max}})^2\,,
\ea
where we have made use of $n^2 (\delta q)^2 = (\Lambda_{max} - \Lambda_{EW})^2$ and have kept the dominant term in the second line
of (\ref{fermion2}).
One can see from (\ref{fermion2}) that $\delta m_{H}^{2}$ becomes very large when $\Lambda_{max} \gg \Lambda_{EW}$, i.e.
$\delta m_{H}^{2} \propto \Lambda_{max}^2 \gg \Lambda_{EW}^2$.  In
particular, if $g_{Y}^{2}$ is still varying when one reaches the Planck scale then it is this scale which provides a physical cutoff. This is the case with three generations where the top quark Yukawa coupling is the dominant one as can be seen from Fig. \ref{fig1}.

If, on the other hand, $\Lambda_{max} \sim \Lambda_{FP} \sim O(\Lambda_{EW})$ as is the case with a heavy fourth generation discussed above, we are faced with a couple of options- and this is where the solution to the Landau pole comes in. Although the cutoff scale can be of O(TeV), the correction $\delta m_{H}^{2} $ also depends on the value of the coupling at that scale. If we stay with the one-loop result in the $\beta$ functions, as we can see from Fig. \ref{fig0}, $\,g_{Y}^{2}$ blows up at the cutoff (Landau pole) and $\delta m_{H}^{2} $ would be out of control even if the cutoff is finite. On the other hand, with the quasi fixed point now being the solution to the Landau pole problem, $\,g_{Y}^{2}$ has a finite value and, as a consequence, $\delta m_{H}^{2} \propto \Lambda_{FP}^2 $. The mass correction coming from the two-loop quartic contribution proportional to $\lambda^2$ can be treated in a similar
fashion with $g_{Y}^{2} \rightarrow \lambda^2$ and $c \sim O((1/16\pi^2)^2)$ yielding a similar conclusion.

One may wonder whether the above argument based on the Higgs-Yukawa sector is sufficient for the statement   $\delta m_{H}^{2} \propto \Lambda_{FP}^2 $ to be correct when we turn on the gauge couplings. However, if the gauge sector also has a fixed point around $\Lambda_{FP}$, it will imply that this might be the true physical cutoff scale of the SM.
As mentioned above, one might have situations in which the gauge sector exhibits a quasi fixed point at the two-loop level e.g. the scenario described in \cite{hunggen}.  Another possibility is the model involving a heavy fourth generation described in \cite{pqchi1}.
As it is emphasized in \cite{pqchi1}, when $\Lambda_{FP}$
is reached, one is no longer justified in evolving the gauge couplings beyond that scale. In fact, a look at Fig. \ref{fig2}
reveals that, as one approaches the quasi fixed point (to be shown subsequently) from below, there is a region very close to it
where bound states and condensates- many of which carrying the SM quantum numbers- get formed \cite{pqchi1} and interact
with the gauge bosons. The naive gauge coupling evolution using the two-loop $\beta$ functions can obviously not be trusted. In fact, it may happen that these extra composite degrees of freedom can lead to a quasi fixed point in the gauge sector i.e. one may have $\beta(g_i)=0$ at the two-loop level, where $g_i$ refers to the three SM gauge couplings.
The same remarks apply to the light fermion Yukawa couplings beside the possibility that they reach a fixed
point close to the ones mentioned above.
For this reason, we will assume from hereon that $\Lambda_{max}$ (or $\Lambda_{FP}$ as discussed below) provides the true
physical cutoff scale.

Under what conditions could the Higgs-Yukawa sector give rise to a TeV-scale physical cutoff? From the above discussion, one can deduce that this
would happen if the couplings in the Higgs-Yukawa sector reach some {\em quasi fixed points} at a TeV scale. 
The next question concerns whether or not such quasi fixed points exist in the SM.

\begin{figure}
\includegraphics[angle=0,width=8.5cm]{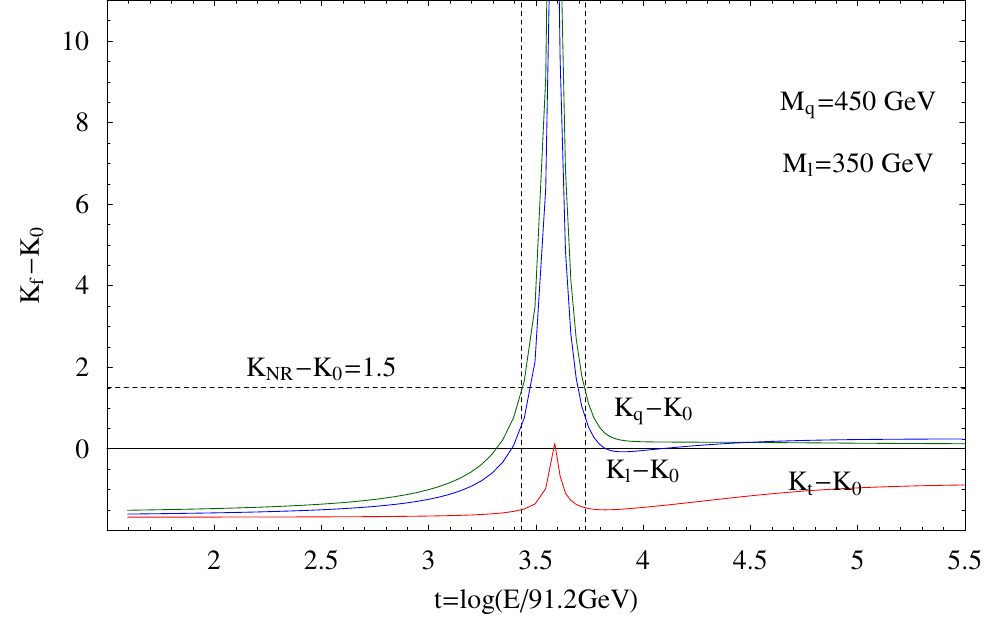}
\caption{$K_f - K_0$ with $ K_f = g_f^3 / 16 \pi \sqrt{\lambda}$ and $K_0=1.68$. The horizontal dotted line indicates 
an estimate of $K_f$ where the non-relativistic 
method is still applicable and the vertical dotted lines enclose the region where a fully relativistic approach is needed.}
\label{fig3}
\end{figure}
In Fig.\ref{fig1}, we show the evolution of the Higgs quartic and top Yukawa couplings at two loop for two initial values
of the Higgs mass: 170 GeV and 500 GeV.
It is amusing to note that quasi fixed points also seem to exist at two loops in the three generation case. However, one can see from
Fig.\ref{fig1} that such a fixed point is either around the Planck scale (heavy Higgs case) or beyond
it (light Higgs case), in which case the cutoff is the Planck scale itself. 
This is the {\em classic hierarchy problem of the SM with three generations}.
The most economical way to lower the cutoff scale  
would be to modify the particle content, e.g. by adding a fourth generation \cite{pqchi1} or by adding
extra chiral doublets. Let us then start with the SM endowed with four generations. Studies 
performed in the past few years have shown that precision data do not exclude 
the existence of the fourth generation \cite{Kribs:2007}. 
Furthermore, if the fourth generation
were to exist, experimental constraints (under a certain assumption) from the Tevatron put a lower bound on the mass at around
$338-385$ GeV \cite{cdf}.

The two-loop renormalization group equations (RGE) for the Higgs-Yukawa sector with four generations are given by
\be \label{RGE}
16 \pi^{2} \frac{dY}{dt} = \beta_Y \,,
\ee
where $Y$ represents the quartic coupling $\lambda$, the Yukawa couplings
$g_t^2$, $g_q^2$, $g_l^2$ for the top quark and the fourth quark and lepton respectively, and the gauge couplings $g_i^2, i=1,2,3$. 
Explicit expressions for $\beta_Y$ up to two loops can be found
in \cite{pqchi1}.

As it was done in \cite{pqchi1}, we first set the $\beta$ functions of
the Higgs-Yukawa sector to be equal to zero to find the fixed points (at two-loop level), namely
\be \label{fixedpoint}
\beta_{Y} |_{{g_{1,2,3}=const.}} = 0, ~~\textrm{for}~~Y= \lambda, g_t^2, g_q^2, g_l^2.
\ee
The roots of (\ref{fixedpoint}) yield the values of the fixed points:
\be \label{fpvalues}
{\lambda}^{*} \approx 17, ~g_{t}^{2*} \approx 31,~g_{q}^{2*} \approx 52, ~g_{l}^{2* }\approx 54
\ee
which correspond to the $\overline{MS}$ masses (using $\overline{m}_H =  v\sqrt{2 \lambda} $ 
and $\overline{m}_f = v \sqrt{g^2_f/2},~v=246$ GeV )
%
$\overline{m}_H^{*} = 1.44~\textrm{TeV}, \overline{m}_t^{*} = 0.97 ~\textrm{TeV}, 
\overline{m}_q^{*} = 1.26 ~\textrm{TeV}, \overline{m}_l^{*} = 1.28 ~\textrm{TeV},$
where the asterisks refer to the values of the masses at the fixed points. Notice that
$(g^2_f, \lambda) / 16 \pi^2$ are typical expansion parameters and, with the fixed point values given above,
these parameters are estimated to be $g^{2*}_t / 16 \pi^2 \approx 0.2,\,g^{2*}_q / 16 \pi^2 
\approx 0.33,\,g^{2*}_q / 16 \pi^2 \approx 0.34$ plus 
$\lambda^{*}/ 16 \pi^2 \approx 0.11$, which are not large. The fixed point values given in (\ref{fpvalues})
are comparable to those found in \cite{Hung:1997rw} but which were used in a different context, that of SU(5) gauge
coupling unification. A remark is in order here: we have neglected the b-quark and $\tau$-lepton Yukawa
couplings in the RGEs but the structure of the $\beta$ functions suggest that they also reach a quasi fixed point \cite{pqchi2}. Also the $\beta$ functions of the gauge sector could reach a quasi fixed point at the two-loop level when additional composite degrees of freedom are included, as we have mentioned above.
Notice the comments concerning the light fermion Yukawa couplings made above.
The energy scales where the quasi fixed points appear which are presented here and in \cite{pqchi1} depend {\em primarily} on the fourth
generation.

By themselves,
the above quasi fixed points {\em do not} tell us about the energy scales where they appear since these values depend mainly on group-theoretical
coefficients which enter the RGEs and on the initial values of the couplings at the electroweak scale. It goes without saying that it is the {\em dynamics} of the SM which would determine 
the {\em values} of these scales. 
Intuitively speaking, one expects that the larger the initial values
of quartic and Yukawa couplings at the electroweak scale are, the ``faster'' they get to the fixed points.
In \cite{pqchi1}, we show that the Higgs-Yukawa sector is almost decoupled from the gauge sector and 
that the quasi fixed points which are the roots of (\ref{fixedpoint}) are affected very little
by the presence of the gauge couplings.

At this point, it is worthwhile to reiterate the following point referred to above. As shown in Fig. (\ref{fig0}), the energy scale where the one-loop Landau singularity appears more or less ``coincides" with that where the two-loop quasi fixed point appears for a heavy 4th generation. In this respect, it is the energy scale $\Lambda_{FP}$ (=$\Lambda_{max}$ here) that is crucial rather than the actual values of the couplings at the quasi fixed point. In fact, if one were to include (unknown) higher orders beyond two loops to the RGE's, it might happen that the actual values of the couplings at $\Lambda_{FP}$ might be lower while preserving the fixed point structure i.e. $\beta=0$. As we referred to earlier, if scale symmetry is restored at $\Lambda_{FP}$ as hinted from the two-loop result, the physics at $\Lambda_{FP}$ or below will not depend on the physics above it.

The energy scales where the fixed points are located can be found by numerically integrating the RGEs (\ref{RGE}). The results
are shown in Fig. \ref{fig2} for two widely separated values of the 4th generation masses. The experimentally 
disallowed case with the smaller 4th generation quark 
mass $M_q=120\, \textrm{GeV}$ is shown only for comparison and to illustrate the naturalness issue discussed below. There are several
implications we would like to present concerning Fig. \ref{fig2}.

1) The quasi fixed points obtained by the RGE evolution agree well with those obtained by setting $\beta_{Y}= 0$. They
are $\overline{m}_H^{\textrm{\tiny{FP}}} = 1.446 ~\textrm{TeV}, \overline{m}_t^{\textrm{\tiny{FP}}} = 0.965 ~\textrm{TeV}, 
\overline{m}_q^{\textrm{\tiny{FP}}} = 1.260 ~\textrm{TeV}, \overline{m}_l^{\textrm{\tiny{FP}}} = 1.282 ~\textrm{TeV}$.

2) The locations in energy scale of the quasi fixed points for the two illustrated examples can be read from Fig. \ref{fig2}. For $M_q=450\, \textrm{GeV},
M_l=350\, \textrm{GeV}$, one has $\Lambda_{FP} \approx 3\,\textrm{TeV}$. For $M_q=120\, \textrm{GeV}, M_l=100\, \textrm{GeV}$, 
one has $\Lambda_{FP} \sim 10^{16}\,\textrm{GeV}$. 
One cannot fail but to
notice that the {\em heavier} the fourth generation is the {\em lower} the fixed point $\Lambda_{FP}$ becomes. As we have
argued in the beginning of the paper, $\Lambda_{FP}$ could be considered to be a {\em physical cutoff}
and that the mass correction to the Higgs scalar is proportional to the square of that cutoff. From this, one can infer
that, not only a light 4th generation such as the $M_q=120\, \textrm{GeV}, M_l=100\, \textrm{GeV}$ case is ruled out by experiment, it
is also ``disfavored'' from a theoretical viewpoint. This
leaves us with a ``heavy'' fourth generation scenario with a TeV-scale physical cutoff scale $\Lambda_{FP}$ \cite{Holdom:2006mr}. 
This, as we claimed above,
might be a possible  solution to the hierarchy problem, in addition to being the solution to the
Landau pole problem present at the one-loop level. 

3) Although our results were obtained with a heavy fourth generation, one can envision a situation in which a fourth generation
with mass around $400-500$ GeV  and endowed with TeV-scale physical cutoff scale $\Lambda_{FP}$ is replaced by several
chiral doublets with {\em lower masses} such as the mirror fermions which are used in the model of electroweak-scale
right-handed neutrinos \cite{hungnu}. In fact, bound states and condensates get formed as one approaches $\Lambda_{FP}$ and
this necessitates a non-perturbative treatment. The appropriate framework for such non-perturbative treatment is to
put the SM on a lattice. A gauge-invariant lattice formulation of the SM is possible only if one introduces {\em mirror fermions}
\cite{montvay}.

4) The ``dips'' in  Fig. \ref{fig2} correspond to a minimal value of $\lambda$ at the 
electroweak scale for which the vacuum stability ($\lambda>0$) is satisfied.
They are located at $\Lambda_{FP} \sim 3\,\textrm{TeV}, 10^{16}\,\textrm{GeV}$ corresponding
to $M_q=450\, \textrm{GeV}, M_l=350\, \textrm{GeV}$ (heavy) and $M_q=120\, \textrm{GeV}, M_l=100\, \textrm{GeV}$ (light) respectively. However,
numerical calculations \cite{pqchi1} have shown that, in order for $\lambda_{dip} \sim 0$ at the dips and to ensure vacuum
stability, one needs to fine-tune the initial value of $\lambda$ to 2 decimal places for $\Lambda_{FP} \sim 3\,\textrm{TeV}$ and to more
than eight decimal places for $\Lambda_{FP} \sim 10^{16}\,\textrm{GeV}$. In the latter case, if one fine-tunes the initial $\lambda$
to less than eight decimal places, it will turn negative and the vacuum will be unstable. It is not so with the ``heavy'' case.
It is amusing to note that the ``light'' fourth generation is not only ruled out by experiment but also theoretically
by the hierarchy and naturalness problems.
This {\em vacuum stability naturalness} issue is deeply linked to the hierarchy problem: A heavy fourth generation might provide a solution to the hierarchy
problem and, at the same time, is devoid of the naturalness problem. 

5) Around the dip and its vicinity i.e.near or at the fixed point, the Yukawa couplings of the fourth generation quarks
and leptons become large and lead to the formation of bound states. This is studied in details in the companion paper 
\cite{pqchi1} and we will summarize our results here. To gain insight into the bound state formation, we start with
the range of $\lambda$ near the fixed point where one can use the Schr\"{o}dinger equation with a Yukawa potential
of the form $V(r) = - \alpha_Y(r) (e^{-m_H(r) r}/ r)$ where $\alpha_Y = \frac{m_1 m_2 } {4\pi v^2} $ with $v=246$ GeV.
The bound state condition of a non-vanishing binding energy translates into the
constraint on $K_f \equiv \frac{\alpha_Y m_f}{m_H}= \frac{g_f^3}{16 \pi \sqrt{\lambda}}$ which is $K_f \geq 2$ and $K_f \geq 1.68$ using a Rayleigh-Ritz
variational technique and a numerical integration respectively. Using the fixed-point values for the fourth
generation and the top quark, we obtain $K_q=1.82$, $K_l=1.92$, and $K_t=0.82$ which imply that fourth-generation bound states
are rather loose and there are no top-quark bound states in this region. As shown in Figs. \ref{fig2},\ref{fig3}, the quartic coupling
decreases rapidly (with the Yukawa couplings being nearly constant) as one moves away from the fixed point value and the 
Yukawa interactions become increasingly long-range. The correlation length $\xi_{H} \sim 1/m_H $ goes from a short-range 
correlation (small $\xi$) to an infinite-range correlation ($\xi= \infty$) at the ``dip'' where one expects condensates
and tight bound states to be formed, e.g. $\langle \bar{Q}_L Q_R \rangle \sim - c \Lambda_{FP}^3$ where $c$ is a constant
which depends on the details of the dynamics. These condensates which contribute to the spontaneous breakdown of the SM
come from extra composite Higgs doublets formed from the quarks and leptons of the fourth generation. (From the
previous discussion, it is not clear whether or not there could be top condensates.) One expects also bound
states with various spins of the form $\bar{Q} Q$, $\bar{L} L$ and even ``leptoquarks'' $\bar{Q}L$ +H.c. to form
due to the strong Yukawa interactions near the ``dip''. These issues will be presented in \cite{pqchi2}.

6) Last but not least, below $\Lambda_{FP}$ (but close to it) dynamical electroweak symmetry breaking occurs \cite{pqchi2} at values of the Yukawa couplings which are smaller than the quasi fixed point values as can be seen from Fig. (\ref{fig2}) at the location of the  ``dip". Above $\Lambda_{FP}$, we conjecture that $\beta=0$ is preserved even (unknown) higher orders are included in the RGE equations and scale symmetry is restored. If that is the case, the physics below $\Lambda_{FP}$ will be independent of what goes on above it. The possibility of the existence of a TeV-scale physical cutoff leads to interesting implications:
(1) The dynamical breaking of the electroweak symmetry below $\Lambda_{FP}$ and (2) The restoration of scale symmetry above $\Lambda_{FP}$ possibly leading to an interesting scale-invariant theory beyond the SM.
\begin{acknowledgments}

This work is supported
in parts by the US Department of Energy under grant No.
DE-FG02-97ER41027.
\end{acknowledgments}


\begin{thebibliography}{99}
\bibitem{models}
The list of references related to these topics is extensive and it is impossible to put it in this note.
\end{thebibliography}
\end{document}